# A compact high-repetition-rate intense THz source for atomic-scale materials characterization

*Léo Guiramand[1], Florant Exertier[2], Alexandre Fahy[1], Anna Martinez[3], Matteo De Tullio[2], Saïd Idlahcen[1], Jonathan Houard[2], Gerald Da Costa[2], Thomas Godin[1], Xavier Ropagnol[4], Domenico Paparo[5], François Blanchard[4], Angela Vella[2,*], and Ammar Hideur[1,*]*

[1]CORIA, UMR6614 CNRS, INSA Université de Rouen Normandie, 76801 Saint Etienne du Rouvray, France

[2]GPM UMR6634 CNRS, INSA Université de Rouen Normandie, 76801 Saint-Etienne du Rouvray, France

[3]Dipartimento di Fisica 'E. Pancini', Universita 'Federico II', Monte S. Angelo, via Cintia, 80126 Napoli, Italy

[4]Ecole de technologie supérieure, Montréal, QC H3C 1K3, Canada

[5]ISASI—Institute of Applied Sciences and Intelligent Systems, Consiglio Nazionale delle Ricerche, via Campi Flegrei 34, 80078 Pozzuoli, Italy

ABSTRACT : High-intensity terahertz (THz) sources combining high repetition rates, strong field amplitudes and long-term stability are key enabling technologies for next-generation ultrafast photonics. However, simultaneously achieving these characteristics in a compact and practical architecture remains a major challenge. Here, we report a high-performance tabletop THz source delivering intense single-cycle pulses at 100 kHz, generated by optical rectification in lithium niobate using an optimized echelon-based tilted-pulse-front scheme and driven by an industrial-grade ytterbium laser. The source achieves optical-to-THz conversion efficiencies exceeding 0.13% at room temperature, peak electric fields of several hundred kV/cm, and excellent long-term stability, establishing a robust platform for high-average-power strong-field THz applications. As a proof of concept, the source was integrated into an atom probe tomography (APT) instrument. APT reconstructs the three-dimensional chemical composition of materials through field-induced ion emission from a sharp nanotip. While THz-assisted ion emission has been shown to improve analytical performance by reducing thermal effects, previous demonstrations were limited to low-repetition-rate THz sources, restricting throughput and accessible analysis volumes. Using an aluminum nanotip, we demonstrate efficient THz-assisted APT at 100 kHz and benchmark its performance against conventional voltage- and laser-assisted operation. The high repetition rate enables rapid large-volume acquisition while improving mass resolving power, reducing thermal tailing, and preserving high-quality three-dimensional reconstructions. Beyond APT, this work establishes a compact and scalable high-repetition-rate THz platform for strong-field physics, nonlinear THz photonics, ultrafast spectroscopy, and material characterization at the nanoscale.

## INTRODUCTION

Due to their unique carrier-envelope phase stability, large ponderomotive energy and low photon energy, intense single-cycle terahertz (THz) pulses offer a wide range of applications based on the manipulation of matter with light[1]. Over the last decade, the development of table-top THz sources has rapidly enabled THz-assisted emission and manipulation of electrons from metallic nanostructures[2–9]. In the THz regime, the emission process is strongly governed by the electric field strength, and the use of nanometric objects such as metallic nanotips enables strong local field enhancement of several V/nm[10]. Although THz-driven electron field emission has been extensively studied in the last decade, THz-induced ion field emission was only recently demonstrated, in 2021, by Vella *et al*[11]. In this athermal process, positive ions are evaporated from a metallic nanotip through the ultrafast lowering of the surface energy barrier driven by the applied THz field. This mechanism has since been applied to atom probe tomography (APT), an analytical microscopy technique that enables 3D reconstruction of needle-shaped nanoscale specimens with near-atomic resolution[12]. Known as THz-assisted APT, this approach offers optimal spatial and chemical resolutions compared to conventional APT instrument where either voltage pulses or laser pulses are applied to the apex of the nanotip. However, to date, all THz-assisted APT demonstrations have relied on intense THz sources based on two-color plasma generation driven by Ti:sapphire laser systems operating at repetition rates limited to a few kHz[13–16]. This intrinsic limitation of the THz source severely restricts acquisition speed and prevents the efficient APT analysis of large volumes containing several million ions. Therefore, the development of intense THz sources operating at significantly higher repetition rates — on the order of hundreds of kHz

— has become critical for advancing THz-assisted APT towards practical applications. Recent progress in high-repetition-rate femtosecond near-infrared ytterbium (Yb) laser technology has opened up new opportunities for high-average-power THz generation. While two-color plasma THz sources operating at high repetition rates have been demonstrated using Yb lasers[17–19], they require pulse post-compression to a few tens of femtoseconds. This leads to complex and costly experimental architectures that limit widespread adoption. For example, in 2021, Buldt *et al.* demonstrated a powerful and broadband THz source with an average power of 640 mW at a repetition rate of 500 kHz, generated from a two-color plasma driven by 37 fs compressed near-infrared pulses[18]. To achieve intense THz pulses, the optical rectification (OR) of sub-picosecond optical pulses in organic or inorganic nonlinear crystals is an attractive alternative. Furthermore, despite a limited field strength compared to plasma sources, OR-based sources can nevertheless achieve high ponderomotive energies thanks to more significant emission of low THz frequencies. At first glance, organic crystals appear particularly efficient due to their large nonlinear coefficients and broadband THz emission. However, their low thermal conductivity limits laser pumping to only a few watts, severely constraining the achievable average THz power[20–23]. In contrast, THz generation by OR in inorganic crystals such as lithium niobate (LN) enables much higher average THz power through the use of pump lasers combining high pulse energy and high repetition rate[24–30]. Ytterbium-laser-driven lithium niobate-based THz sources exhibit outstanding performance in terms of average THz power, together with relatively high optical-to-THz conversion efficiencies of around $10^{-3}$ – $10^{-2}$ [31]. As a result, this scheme has recently demonstrated impressive performance, with 445 mW average THz power at 100 kHz[29]. However, reaching such a level requires cryogenic cooling of the LN crystal using liquid nitrogen, resulting in a conversion efficiency of 0.11% at an average pump power of 500 W. In addition, this THz source relies on a grating-based tilted pulse-front (TPF) configuration, which is highly sensitive to alignment and susceptible to angular dispersion effects[32,33]. Ytterbium-laser-driven optical rectification in lithium niobate is a promising approach for producing high-repetition-rate (≥100 kHz) single-cycle THz pulses compatible with strong-field physics applications and requiring high detection sensitivity, such as THz-assisted APT. Nevertheless, the practical implementation of these applications calls for further breakthroughs to improve the efficiency of such THz sources while simplifying their architectures.

In this work, we combined several solutions to develop a high-performance tabletop THz source based on ytterbium-laser driven optical rectification in lithium niobate. As a first key parameter, we implemented a discrete tilt of the pump laser wavefront using a reflective gold coated echelon mirror[27,33–35]. This approach is comparatively straightforward to align and promotes higher optical-to-THz conversion efficiency by avoiding the dispersive effects inherent to continuous TPF configurations[33]. A second critical parameter that we explored was the use of relatively long, narrow-band pump pulses to favor low-frequency THz generation and suppress spectral components above 3–4 THz, which are strongly absorbed in lithium niobate. Optimizing these parameters enabled the realization of an intense and highly stable THz source operating at a repetition rate of 100 kHz with an optical-to-THz conversion efficiency exceeding 0.13%. We subsequently coupled this high-performance THz source with a custom-built APT system. Using this platform, we demonstrated high-repetition-rate THz-assisted APT operation on a metallic specimen for the first time. Direct comparison with conventional laser-assisted APT confirmed the strong potential of high-repetition-rate THz-assisted APT for the precise, atomic-scale analysis of materials.

## RESULTS

The schematic overview of the high-repetition rate THz-assisted APT is shown in **Figure 1**, as well as its basic operation. An ultrafast laser is used to generate single-cycle THz pulses, which are then focused onto a nanotip-shaped sample inserted inside an ultrahigh vacuum chamber. These pulses trigger the emission of atoms from the positively polarized sample. A time and position-sensitive detector measures the flight time and impact position. The flight time is used for the chemical identification of ions via time-of-flight (TOF) mass spectrometry. From the impact position of the ions, a 3D image of the sample is reconstructed using a back-projection law. Our high-repetition-rate THz-APT prototype then consists of three main stages: a high-intensity single-cycle THz source, a THz-waveform tailoring stage, and an APT-system. These are described briefly in the following sections and more details can be found in the *Materials and Methods* section.

### High-repetition rate THz generation and wavefront tailoring

An Yb-doped solid-state laser system (Carbide, Light Conversion) is used to drive THz emission (pump), THz electro-optic (EO) sampling and laser-assisted APT experiments (probe). The pump beam is used to generate THz radiation through optical rectification in a lithium niobate crystal at room temperature. To satisfy the phase-matching conditions while avoiding dispersion effects, a reflective stair-step echelon mirror is used to tilt the pump beam wavefront in a discrete manner. To optimize the interaction area inside the nonlinear crystal and the wavefront tilt angle, a cylindrical lens is used to shape the laser beam into a line-like area. The THz radiation emitted by the LN crystal is collected by an off-axis ellipsoidal mirror (OAEM) and characterized using a calibrated thermopile THz detector (OPHIR 3A-P-THz). The first characterizations conducted in this work was the analysis of the THz emission performance while varying both the pump repetition rate and energy.

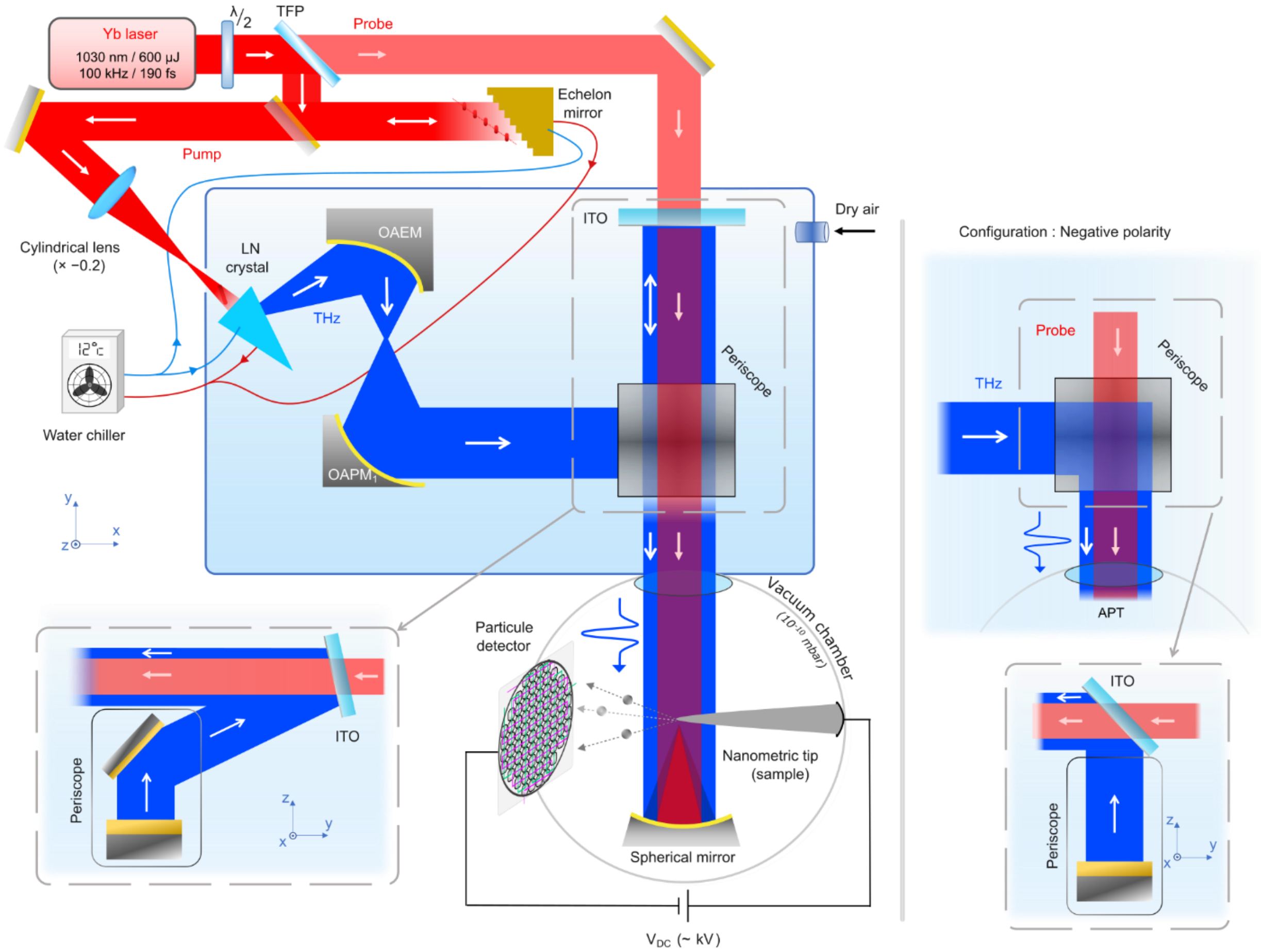

**Figure 1**: Overview of the THz-assisted atom probe tomography setup. TFP: thin film polarizer, λ/2: half- wave plate, OAEM: off-axis elliptic mirror, OAPM: off-axis parabolic mirror, ITO: indium tin oxide.

As summarized in **Figure 2**(a), the optical-to-THz conversion efficiency remains unchanged at fixed pump energy over an extended operation, evidencing efficient thermal management of the water-cooled reflective echelon and LN crystal[25,29]. This result is particularly significant at high repetition rates, where heat accumulation can induce temperature-dependent phase-matching detuning and consequently reduce THz generation efficiency[36]. At 100 kHz repetition rate, the source generates 80 mW average THz power with an overall efficiency of more than 0.13%. This performance demonstrates successful repetition-rate scaling of the echelon-based TPF scheme while maintaining a conversion efficiency comparable to state-of-the-art results obtained at lower repetition[35].

In addition, the echelon architecture provides a substantially simpler optical implementation than conventional grating-based TPF schemes, offering improved robustness and a clear pathway toward industrial-grade THz systems. To collimate the THz beam after the OAEM, an off-axis parabolic mirror ($OAPM_1$) is used. For THz-assisted APT experiments, a linearly polarized THz beam whose polarization is aligned along the nanotip axis (in the x–y plane) is required to optimize field enhancement at the tip apex[11]. As the generated THz radiation is polarized along the c-axis of the lithium niobate crystal (i.e., along the z-axis in **Figure 1**)[37], its polarization must be rotated by 90° using the periscope shown in **Figure 1**. The THz pulses generated through OR exhibit a near-single-cycle temporal profile characterized by an asymmetric electric-field waveform, commonly referred to as its polarity[38]. Precise control of this polarity is critical for THz-assisted APT, as it determines the direction of the transient electric field at the nanotip apex. Accordingly, two polarity configurations were investigated: a positive polarity, for which the peak THz field is directed toward the apex, and a negative polarity, for which it is directed away from the apex. Switching between these configurations is achieved by introducing an additional reflection in the THz beam path, resulting in a $\pi$ phase shift of the THz pulse envelope. Finally, an indium tin oxide (ITO)-coated glass plate is used to combine the probe laser beam and the THz beam colinearly. Immediately before entering the APT chamber, the THz beam delivers up to 43 mW of average power. The long-term stability of the source was evaluated for an average THz power of 27.43 mW, as shown in **Figure 2** (b). Following a 13 min thermal equilibration period, the THz power was monitored continuously for 3.5 h. The measured fluctuations exhibited a standard deviation of only ±0.25% relative to the mean power, demonstrating exceptional source stability. Such performance is a key requirement for reproducible, high-throughput THz-assisted APT.

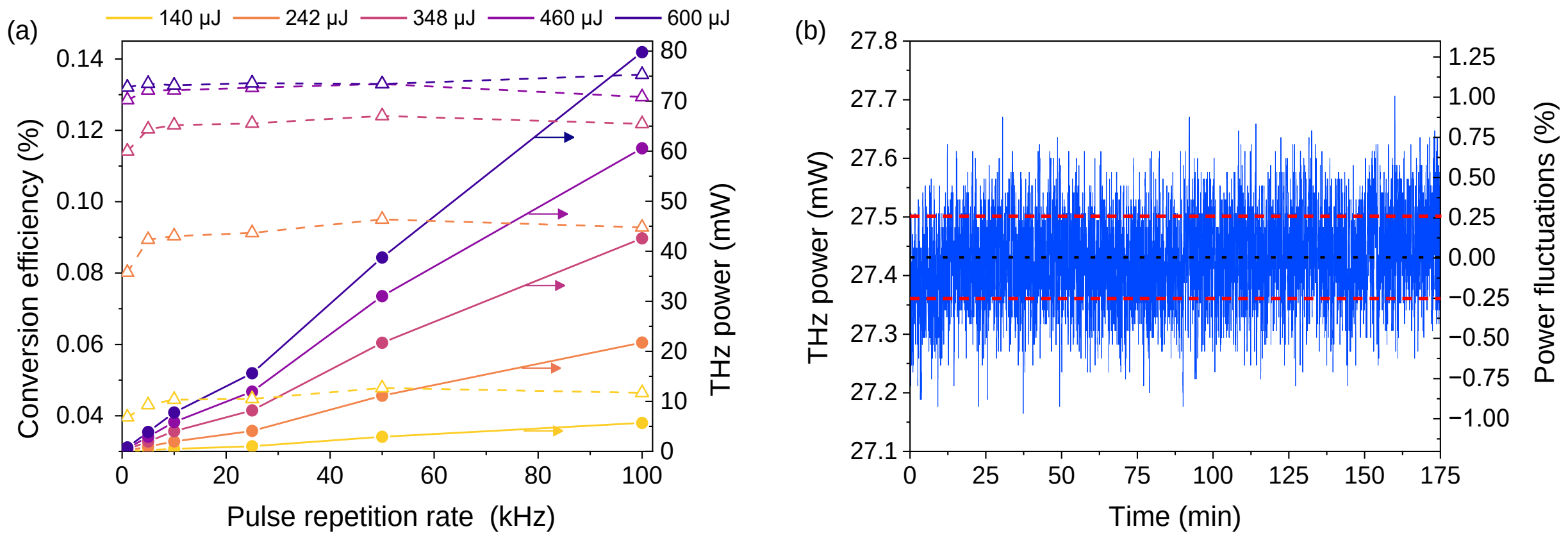

**Figure 2**: THz power characterization. (a) Average THz power measured at the focus of the OAEM and corresponding optical-to-THz conversion efficiency as a function of pulse repetition rate for different pump levels. (b) Long-term THz power stability over 3.5 h. measured immediately before the APT chamber in the positive-polarity configuration and for a THz power of 27.43 mW. The black dashed line indicates the mean THz power (27.43 mW), while the red dashed lines denote one standard deviation (±0.07 mW).

The temporal waveform of the THz pulses just before the APT chamber was measured using EO sampling with a thin x-cut LN crystal deposited on a 1 mm-thick z-cut LN substrate (see *Supplementary Material,* section S3). **Figure 3**(a) and (b) show the pulse temporal profile and its associated amplitude spectrum, respectively, measured at a maximum pump energy of 600 $\mu$J and a repetition rate of 100 kHz. The pulse exhibits a single-cycle behavior with an asymmetric polarity, and a duration of approximately 0.82 ps, as estimated from the Hilbert transform. Its frequency distribution is centered at 0.85 THz with a full width at half maximum (FWHM) of 0.9 THz. A detailed study of the temporal shape and spectrum of the THz pulse under various pumping conditions is presented in the *Supplementary Material,* section S3. The THz peak electric field can be estimated from EO sampling measurements using two balanced photodiode signals, considering the electro-optical coefficients of the detection crystal[39]. The inferred peak amplitude of the THz field is approximately 250 kV/cm. We also compared the values obtained using this method to those estimated using a method based on measuring the duration, energy, and spot size of the THz pulse[40]. The latter method yields a slightly higher field value of 266 kV/cm, which can be explained by the fact that the method based on EO modulation underestimates the field strength (see *Supplementary Material*, section S4)[41]. We attribute this difference mainly to a convolution effect of the THz pulse with the relatively long optical probe pulse (190 fs duration, FWHM). This results in a smoothing effect on the measured THz wavefront, leading to an underestimation of the peak electric field[42].

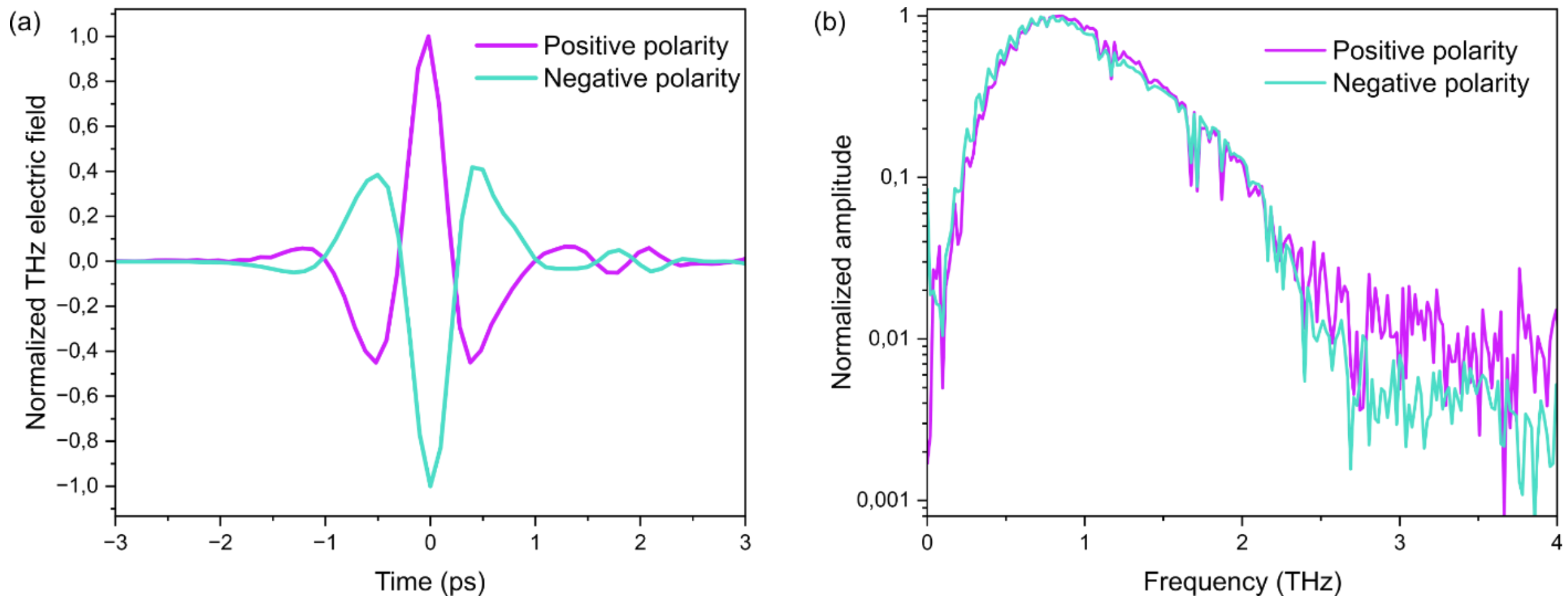

**Figure 3**: THz pulses characterization. (a) Temporal evolution of the THz electric field, acquired by EO sampling, in the case of positive and negative polarity. (b) Corresponding amplitude spectra.

To determine the absolute polarity of the pulses propagating toward the APT in the two proposed configurations, THz-assisted electron emission can be characterized in the time domain. The carrier-envelope phase stability of single-cycle THz pulses makes electron emission highly sensitive to the instantaneous THz-field phase and polarity[9]. The experiment performed in this work is based on laser-assisted electron field emission and is described in detail in the *Supplementary Material*, Section S5[10,43]. This pump-probe experiment also provides information on the temporal wavefront of the THz pulse, which is enhanced by the antenna effect arising from its interaction with the nanotip sample. The experimental setup is essentially identical to that used for the APT experiments described below. In this case, however, we studied electron emission from a pure aluminium (Al) nanotip illuminated by a probe laser pulse and a THz pulse. The THz field was strongly attenuated by five silicon (Si) wafers in order to prevent any electron emission induced solely by the THz contribution. This attenuation does not modify the temporal shape or spectrum of the THz pulses (see *Supplementary Material,* section S5). The electron flux is measured using a particle detector coupled to a lock-in amplifier. The voltage measured by the lock-in amplifier is proportional to the electron flux and can be related to an equivalent static voltage applied to the tip. This requires a calibration process, which is described in detail in the *Supplementary Material,* section S5. The nanotip is negatively biased to -300 V and illuminated by probe laser pulses to trigger electron emission via photoemission, with a yield below 10%. The peak intensity of the probe laser at the nanotip apex is approximately 300 GW/cm$^2$. The THz pulses are simultaneously coupled to the nanotip and contribute to the electron emission process. Given the high electric-field strength and low frequency of the THz pulses used here, the Keldysh parameter is expected to be well below 1, indicating a field-emission regime[44]. In this regime, electron tunneling occurs on a sub-cycle timescale, making THz-assisted electron emission effectively quasi-instantaneous[5].

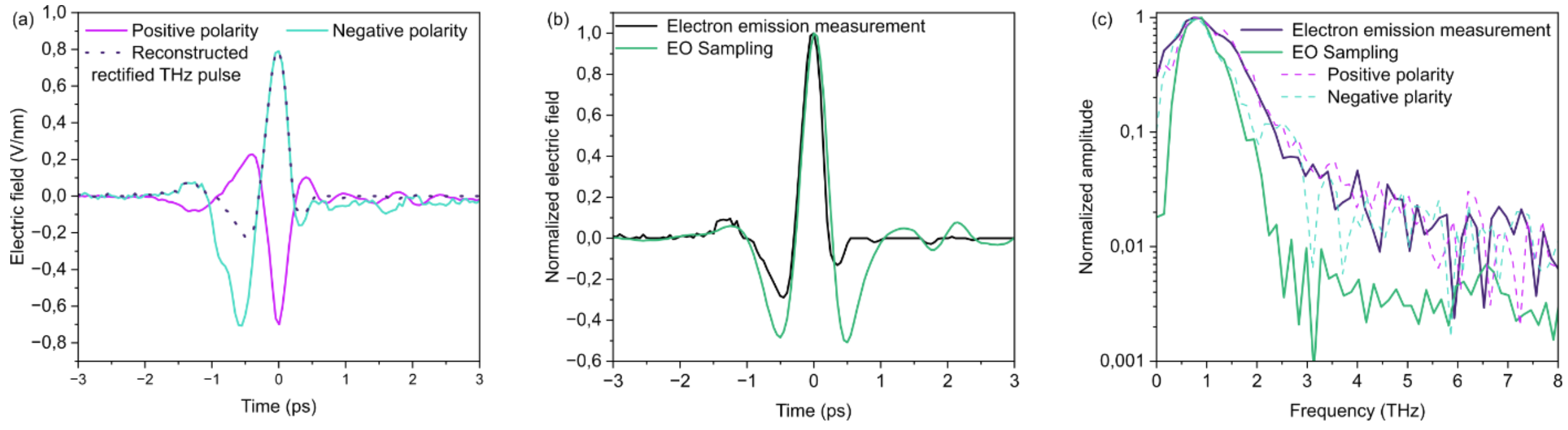

**Figure 4**: Results of a laser-assisted electron field emission experiment. (a) THz electric field at the nanotip apex, inferred from the electron flux measured for different delays between the laser and THz pulses (attenuated by 5 Si wafers). (b) Comparison of the rectified THz field reconstructed from the laser-assisted electron emission experiment with the waveform of the THz pulse obtained by EO sampling. (c) Comparison of the corresponding amplitude spectra.

Consequently, no additional temporal response function is required for the reconstruction of the THz electric field from the measured electron flux. Varying the delay between the IR laser pulse and the THz pulse allows the instantaneous contribution of the THz field to electron emission to be evaluated. Owing to the low electric field strength of the attenuated THz pulse, when the THz pulse arrives before the probe laser pulse (time delay < -1 ps), the emission is dominated by the laser contribution. When the THz pulse temporally overlaps with the probe pulse, the THz field modulates the electron flux: the emission increases in the presence of an additional negative field and decreases in the presence of an additional positive field. As this flux modulation can be related to an equivalent static bias field, the THz field at the nanotip apex position is extracted for both polarity configurations, as shown in **Figure 4**(a). Furthermore, considering only the contribution of the positive electric field in both configurations - which is responsible for the increase in electron emission flux - allows us to reconstruct an equivalent rectified THz pulse (dotted curve in **Figure**

**4**(a)). This is achieved by inverting the electric field corresponding to the positive polarity and combining it with the electric field corresponding to the negative polarity.

**Figure 4** (b) compares the reconstructed rectified THz waveform with the waveform measured by EO sampling, and the corresponding spectra are shown in **Figure 4**(c). Although the two techniques yield consistent temporal waveforms, the reconstructed pulse exhibits a substantially broader spectrum, extending from 0.15 to 1.6 THz. The broader spectral content inferred from the laser-assisted electron field-emission measurements is further supported by considering separately the spectra of the THz electric-field signals measured for the two polarities, without combining the two measurements (dotted curves in Fig. 4(c)). The narrower EO spectrum results from the finite detection bandwidth imposed by the EO crystal, which attenuates the high-frequency components of the THz field. In this pump-probe experiment, the equivalent rectified THz field - attenuated by a factor of 5.88 due to the insertion of five silicon wafers - reaches a maximum field strength of 1.2 V/nm. Without silicon wafers, the THz pulse is therefore amplified by the nanotip to an estimated peak field of approximately 7 V/nm at the apex. Considering the estimated THz peak electric field before the APT chamber (170 kV/cm) and the transmittance of the diamond entrance window (approximately 84%), the field enhancement factor at the nanotip apex is estimated to approximately 500 (see *Supplementary Material,* section S5). This value is of the same order of magnitude as the enhancement factors obtained from simulations of the electric-field distribution at the apex of nanotips[10]. An independent experimental estimate of the enhancement factor was also obtained from electron-emission blocking-voltage measurements, as described in detail in the *Supplementary Material*, Section S5. Consequently, the electric field strength reached at the nanotip apex is sufficient to enable THz-assisted ion emission, as will be demonstrated below.

**Atom probe tomography results and discussion**

In conventional APT instrument, either voltage or laser pulses are applied to the apex of the nanotip-shaped sample in addition to a standing DC bias to overcome the field-evaporation energy barrier. Previous work has shown that THz pulses initiate field evaporation by increasing the local electric field at the tip apex rather than by heating, in contrast to conventional laser-assisted instruments[11]. As a results, thermal artifacts are reduced, leading to improved signal-to-noise ratio (SNR), mass resolving power (MRP) and spatial resolution. By combining high THz field strengths with 100 kHz operation, the source developed in this work enables a two-order-of-magnitude increase in the operating repetition rate of THz-assisted APT compared with previous demonstrations. This performance enables high-throughput APT analysis at repetition rates comparable to those of commercial instruments, while preserving the advantages of THz-driven field evaporation and its low thermal load. This represents an important step toward practical and scalable THz-enabled nanoscale characterization

To validate the performance of this novel high-repetition-rate THz-assisted APT instrument, pure Al specimens were analyzed and compared with conventional APT measurements. Al was deliberately selected as a reference material because its single-element composition and well-defined atomic structure provide well-controlled conditions for evaluating the spectral and spatial performance of the APT instrument [45]. APT measurements were thus successively performed on a single specimen needle with an initial radius of approximately 50 nm, using the following sources in the following order: (*i*) IR laser pulses, (*ii*) negative THz-pulses, (*iii*) positive THz-pulses (first round), (*iv*) UV laser-pulses, (*v*) positive THz-pulses (second round). Additional measurements were carried out on a separate specimen with a similar initial radius of approximately 50 nm to compare the field evaporation behavior with that observed using a commercially available LEAP 5000 XS instrument (CAMECA Instruments), which operates in voltage-pulsing mode.

First, we studied the effect of THz pulse polarity on the field evaporation behavior, comparing the results with those obtained using IR pulses. In this case, the analyses were performed in the same APT system and on the same specimen. **Figure 5**(a) shows the mass spectra acquired using different pulse types. It should be noted that each of the three datasets corresponds to $4.10^5$ ions selected from a total of ~ $2.10^6$ detected ions acquired successively on the same nanotip; therefore, the shape and radius of the nanotip are considered as unchanged (see *Supplementary Material* , section S.7). In the mass spectra, whilst thermal tailing is observed for the $Al^+$ peak (27 atomic mass unit (amu)) in the IR data, no such thermally induced bias is observed in the positive THz data. This illustrates the reduced thermal effects as previously demonstrated by Vella et al.[11], as well as the improved MRP (see *Supplementary Material*-Table S2). When the THz polarity is switched from positive to negative, the $Al^{2+}$ peak splits into two distinct peaks. However, the $Al^+$ peak shows two overlapping peaks, forming a double peak. This is due to the two positive lobes of the THz pulse in the negative polarity, as illustrated in **Figure 3**(a). Each positive lobe of the THz pulse has the potential to trigger ion emission. The time delay in the TOF of ions emitted at the first or second THz maximum increases due to the ions being accelerated and/or decelerated by the THz field in the region close to the sample apex. This deceleration dynamic has already been studied by M. Karam et al.[14], in the case of $LaB_6$ samples analyzed in THz-APT at 1 kHz repetition rate.

The spatial distribution of ions detected during atom probe analyses is often represented as a two-dimensional (2D) map, wherein the positions of impacts on the detector surface are histogrammed. The color coding accentuates regions of low atomic density in blue and regions of high density in yellow. It has been established that the local atomic density is inversely proportional to the magnification in the region of atomic origin. Therefore, such maps provide insights into local variations in curvature and shape of the specimen. The ion-detector impact maps from the first round of analyses are shown in **Figure 5**(b). As illustrated by the maps obtained using THz pulses, a clear structure of poles and zone lines is observed, which is a typical feature of multilayer field-desorption patterns. In the case of a crystalline sample such as aluminum, the conventional crystal structure is interrupted by the hemispherical shape of the tip. This gives rise to regular structures of concentric atomic rings, which are known as poles. The center of the <hkl> pole is defined as the point at which the atomic planes (hkl) intersect with the hemispherical surface. Clearly, the local field exhibits a decrease in magnitude at the center of the pole when compared to its periphery. Consequently, the number of evaporated atoms is reduced at the center. It can thus be concluded that poles are concomitant with regions of lower density on the maps presented in **Figure 5**(b). Using the IR laser elevates the specimen temperature, thereby affecting the entire pattern. Certain poles and zone lines become indiscernible, whilst the dimensions and configuration of the remaining crystallographic features may substantially alter. The increase in density around the Al <002> pole, as illustrated in **Figure 5**(b) for IR analysis, indicates that the global shape of the specimen changes. Furthermore, using the impact position of the ions on the detector and a back-projection algorithm, it is possible to reconstruct the nanotip sample in 3D with the original position of each ion. The reconstructed volumes of the samples under different excitations, with the atomic plane alignments corresponding to the atomic direction <113>, are shown in **Figure 5**(c). The 3D reconstruction of the evaporated volume reveals well-resolved atomic planes for the THz analyses, which were indiscernible in the case of the IR analysis. The spatial resolution values for the THz analyses are presented in the *Supplementary Material* (Table S3). We have also explored the impact of THz power on the APT performances. Our findings show that reducing the THz power does not alter the $Al^+$ peak, confirming that THz-induced heating remains negligible even at high repetition rate; however, at low THz field, increased background noise reveals continuous evaporation, see *Supplementary Materials.* For non-HV-pulsed experiments, the field contribution of the pulse can be expressed in terms of an equivalent pulse fraction: reducing the pulse intensity is equivalent to reducing the pulse fraction in voltage-pulsed experiments. When the pulse contribution to the electric field becomes too small, the DC voltage must be increased to maintain a constant detection rate, which can promote preferential evaporation and alter the measured composition[46,47].

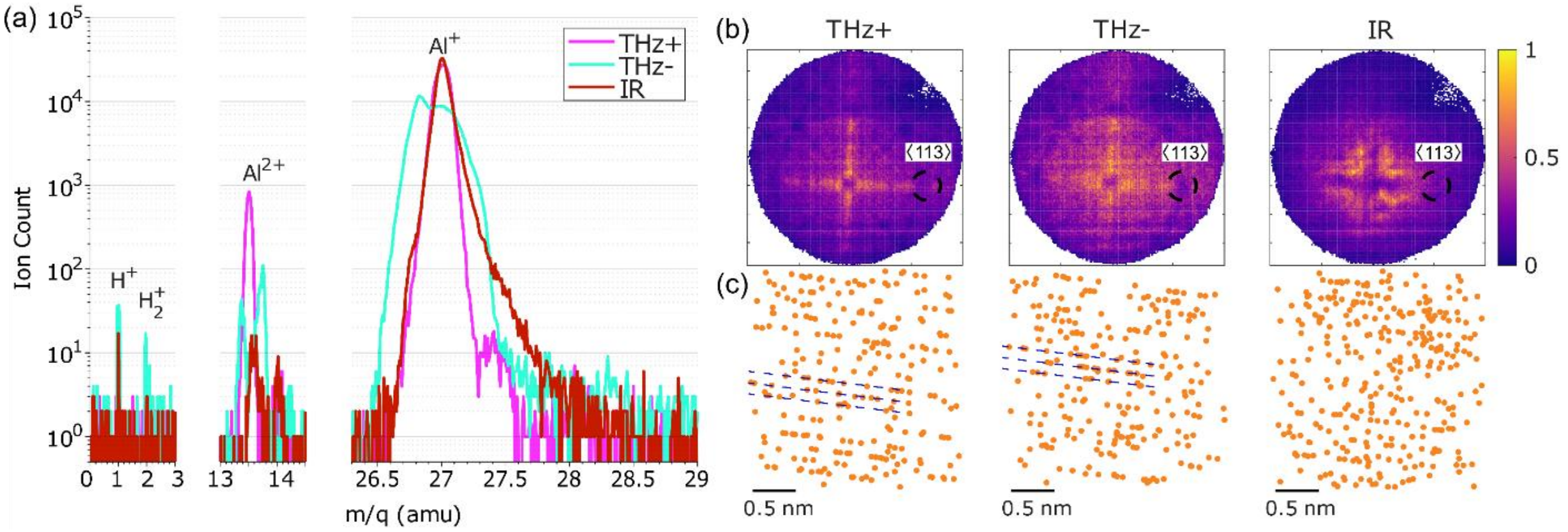

**Figure 5**: Comparison of APT datasets of $4 \cdot 10^5$ ions each acquired from an aluminum sample using positive THz pulses (100% laser power, $V_{DC}$ ~2.3 kV), negative THz pulses (100% laser power, $V_{DC}$ ~2.6 kV) and IR laser pulses ($V_{DC}$ ~2.4 kV). All datasets are obtained with a sample temperature of ~40K, a detection rate of ~ 0.005 ion/pulse and a pulse repetition rate of 100 kHz. (a) Mass spectra, (b) APT detector hit maps, and (c) 3D representation of the atomic plane alignments corresponding to the atomic direction <113> ; blue dashed lines are guides to the eye.

In order to compare the performances obtained using positive THz pulses with that of other pulsing methods used in conventional atom probes, the mass spectra obtained with THz, UV laser and electrical pulses are presented together in **Figure 6**(a). All data sets were obtained with a pulse repetition rate of 100 kHz and a detection rate of approximately 0.005 ion/pulse, maintained constant throughout the entire acquisition by adjusting the DC voltage. The UV and voltage pulsing (VP) samples are analyzed at 30 K, the THz sample at ~ 40 K, the lowest temperature that can be achieved on the THz-APT device. The UV laser energy is 3 nJ, the lowest energy that allowed working with a fixed detection rate of 0.005 ion/pulse without observing continuous evaporation. The contribution of the electrical pulses is set at 20% of the evaporation field. This value is typical for APT analysis of Al using electrical pulses. With THz pulses, the Al ion peaks are very narrow and do not exhibit tails on the right side, unlike what is observed with UV pulses, due to the thermal effect, or with electrical pulses, due to the energy deficit of the ions. It follows that the use of THz pulses improves the mass resolution of the instrument compared to the commercial atom probes.

Furthermore, using THz pulses allows to detect molecular ions such as $AlH^{+}$ or $AlH^{+2}$ , while hydrogen ions or molecules are greatly reduced. However, hydrides are likely to be reduced and might not be visible in the mass spectrum when using a UV laser or high-voltage pulses due to the long tail of the $Al^{+}$ peak. In addition, the presence of more atomic and molecular hydrogen indicates that the electric field of the THz pulses promotes the evaporation of aluminum hydrides. The ion detector impact maps show very well-resolved crystallographic poles and zone lines for all three analyses (see **Figure 6**). In particular, despite the higher fixed sample temperature for the THz analysis, the poles and zone lines remain clearly visible.

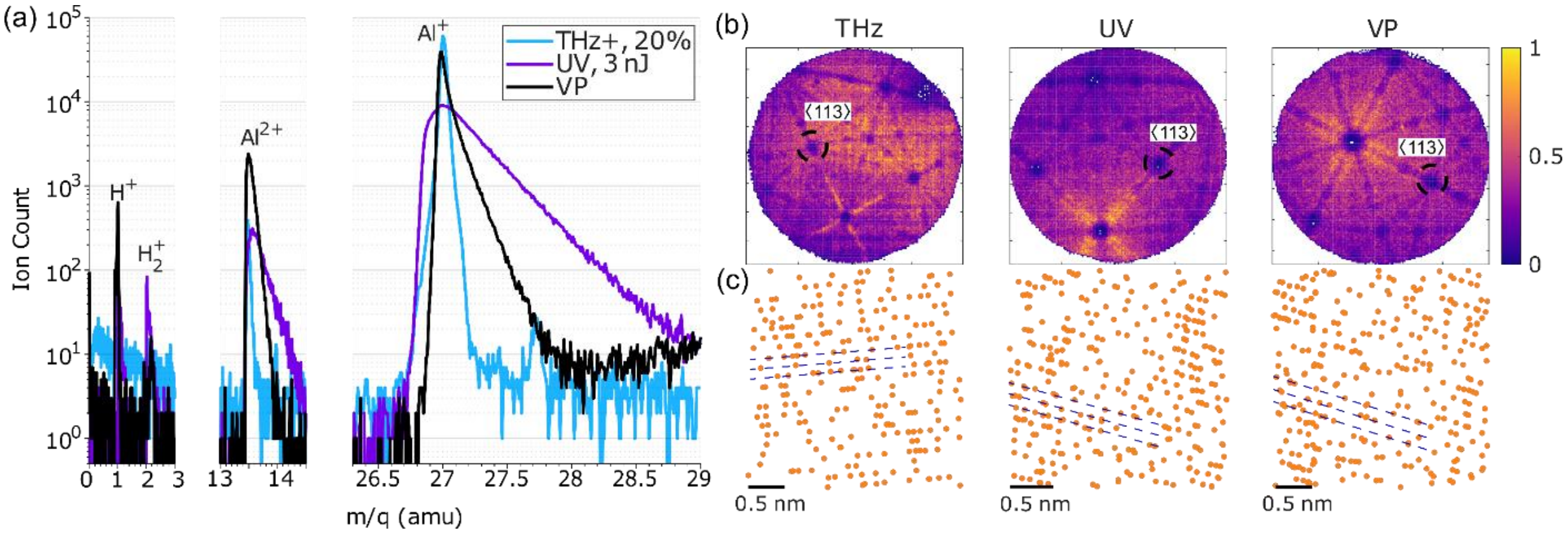

**Figure 6:** Comparison of APT data acquired using positive THz pulses, UV laser pulses and VP on three different nanotips. All datasets are obtained with a detection rate of ~ 0.005 ion/pulse and a pulse repetition rate of 100 kHz. The UV and VP samples are analyzed at 30 K, the THz sample at ~40 K. (a) Mass spectra, (b) APT detector hit maps and (c) 3D representation of the atomic plane alignments corresponding to the atomic direction <113> ; blue dashed lines are guides to the eye.

This proves once again that the heating induced by THz pulses at a high repetition rate is negligible. The atomic planes reconstructed in 3D for the <113> direction is shown in **Figure 6**(c). These planes are well-resolved for all three analysis conditions, but the spatial resolution is better when THz is used. Therefore, using THz pulses improves the spatial resolution of the instrument compared to the commercial atom probe. The MRP and spatial resolutions are detailed in the Supplementary Material, section S7.

## DISCUSSION AND CONCLUSION

By combining a femtosecond Yb laser with an optimized echelon-based TPF configuration, we demonstrate a state-of-the-art high-repetition-rate THz source delivering peak electric fields of several hundred kV.cm$^{-1}$ at 100 kHz while maintaining an overall optical-to-THz conversion efficiency exceeding 0.13% at room temperature. The source combines a compact, industrially compatible architecture with excellent long-term stability, demonstrating that strong-field THz generation can be efficiently scaled to high average powers and repetition rates.

As a proof of concept, the source was integrated into an atom probe tomography (APT) instrument, enabling THz-assisted operation at a repetition rate two orders of magnitude higher than previously reported, enabling routine APT analysis of large volume with ion detection rate up to $10^7$ ions per hour. Analysis of aluminum specimens demonstrated high-throughput operation together with improved mass resolving power, reduced thermal tailing, and high-quality three-dimensional reconstructions. In the near future, we aim to extend this new technology to more complex and challenging materials, including high-entropy metal superalloys, semiconductors, and insulating materials such as silica. This will require further advances in needle-shaped specimen preparation and, importantly, in THz-source technology to achieve higher apex fields, paving the way towards broader and industrially relevant applications. Cryogenic cooling of the LN crystal could be explored as a means of reducing THz absorption and further improving the conversion efficiency.

Beyond THz-assisted APT, this work establishes a robust platform for strong-field THz science requiring the simultaneous combination of high field strength, high repetition rate, and long-term stability. The compact echelon-based architecture provides a practical route toward scalable, industrial-grade THz systems and opens opportunities for applications in ultrafast spectroscopy, nonlinear THz photonics, strong-field electron and ion emission, and advanced nanoscale characterization.

## MATERIALS AND METHODS

### THz source

The THz source that has been developed in this work is driven by a Yb-doped solid-state laser system (Light Conversion, CARBIDE). This industrial-grade laser contributes significantly to the simplicity and compactness of the generation system and ensures stable laser emission over time. The laser delivers 190 fs pulses centered at 1030 nm with pulse energies of up to 600 μJ at repetition rates as high as 100 kHz, corresponding to a maximum average power of 60 W. For low-frequency (< 5THz) emission, the relatively long pulse duration maximizes THz generation by limiting the dispersion of the laser group delay in the LN crystal[48]. At the laser output, a half-wave plate followed by a thin-film polarizer (TFP) splits the beam into pump and probe arms. More than 95% of the laser power is reflected by the TFP and directed to the THz generation stage, while the transmitted fraction serves as the probe beam for electro-optic (EO) sampling or laser-assisted APT measurements. The pump beam illuminates the echelon mirror at near-normal incidence. Reflection from the stepped surface introduces a controlled time delay between adjacent beamlets, producing a discretely tilted pulse front. The echelon mirror has a 20 × 20 $mm^2$ square aperture with a step width (*W*) of 175 μm and a height (*H*) of 80 μm. It is fabricated from aluminum and coated with a gold film capped by a thin $SiO_2$ protective layer, resulting in a reflectivity greater than 80% at 1030 nm, significantly improving the pump-energy handling compared with similar configurations limited by the echelon reflectivity[27,35] . The incident pump beam exhibits a Gaussian intensity profile with $1/e^2$ diameters of 4.4 × 3.6 $mm^2$, corresponding to a peak fluence of 4.8 mJ $cm^{-2}$. To satisfy the phase-matching condition in the LN crystal between the pump pulse and the emitted THz pulse, a tilted image from the echelon mirror is made in the *x-y* plan with a magnification factor *M* given by[34]:

$$M = \frac{2 \times H}{\tan(\gamma) \times W \times n_{LN}^{gr}} \tag{1}$$

where $\gamma$ is the final tilt angle to be achieved in the crystal LN crystal and $n^{gr}{}_{LN}$ is its refractive index at the pump wavelength (2.2 at 1030 nm).

The demagnification factor $M$ is approximately 0.2, achieved using a cylindrical lens with a focal length of 100 *mm*. The generation crystal is prism-shaped with a height of 10 mm, cut at an angle of 63° and composed of stoichiometric LN doped with 1 % of MgO. When placed on the image plane to yield a final tilted angle of 63° inside it, the incident pump beam has a diameter (taken at $1/e^2$) of 0.77 mm × 2.9 mm and each discrete beam have a width of 35 $\mu$m. With an average power of several tens of W, the mirror echelon and LN crystal can heat up, causing thermal issues. These two elements are therefore cooled using a water chiller set to a temperature of 12°C. The generated THz beam exhibits a slight angular divergence[27]. To efficiently collect the generated THz radiation, an off-axis ellipsoidal mirror (OAEM) with dual focal points is used. Set at a distance of approximately 84 mm from the LN crystal, it refocuses the THz beam at a distance of 33 mm. For APT experiments, a collimated THz beam with a diameter of less than 25 mm is required due to the limitation imposed by the ultra-high vacuum chamber entrance window. Therefore, an off-axis parabolic mirror ($OAPM_1$) with a diameter of 25 mm and a focal length of 25 mm is set at 58 mm from the OAEM. A periscope is then used to rotate the THz polarization by 90°. Furthermore, the THz pulse polarity is controlled by adding a reflection in the path of the THz beam. The positive polarity configuration is obtained by using two flat metal mirrors in the periscope, followed by an ITO-coated glass plate to reflect the THz beam towards the APT chamber. Note that the second mirror reflects the THz beam at a small upward angle without altering its polarization direction (see *Supplementary Material,* section S2). The negative polarity configuration is achieved by using an ITO-coated glass plate in the periscope to reflect the THz beam directly towards the APT. Finally, it should be note that, to limit THz absorption related to air humidity, the THz source is placed in an environment with dry air at a relative humidity stabilized at ~ 10%.

**Sample preparation**

Pure aluminum (Al) samples are prepared using a standard electropolishing procedure by etching Al wire using a perchloric acid solution. This results in a final radius at the apex of ~ 50 nm[49].

**APT measurements**

For APT analyses using THz and IR laser pulses, the setup consists of a custom-built ultra-high vacuum (UHV) chamber operating at a typical pressure of ~ $10^{-10}$ mbar. The THz and IR beams enter the chamber through a 25 mm diameter, 500 μm thick diamond window. These beams are then focused on the needle-shaped sample using a spherical metallic mirror with a diameter of 25 mm and a focal length of 25 mm. Inside the UHV chamber, the sample, which has a radius of ~50 nm, is cooled to a cryogenic temperature (typically between 40 K and 80 K) and subjected to a high DC voltage of up to 15 kV to trigger field ionization of the atoms on the surface of the sample. The ions are propelled onto a bespoke position sensitive detector (PSD) that measures the ion impact coordinates as well as their individual time of flight (TOF). The TOF allows for the identification of the ion's mass-to-charge ratio, i.e. their chemical identity (with some limitations). The impact coordinates and the orders of arrival are used to calculate the initial X, Y and Z coordinates of the ions. This provides a 3D map of the sample before field evaporation. The PSD used in this study is a delay line detector with enhanced multihit capabilities. It consists of an 8 cm conventional Roentdek delay line detector, followed by a pair of microchannel plates (MCPs) arranged in a chevron configuration. The signal emitted at each end of the two pairs of wires is digitized using PXI DC 271 (or Keysight U1051A) fast digitization cards. The ion time-of-flight signals are taken at the output of the MCPs, and the positioning is calculated using the signals from the delay line outputs. A deconvolution algorithm is then employed to further improve the PSD's multihit capabilities.

The APT analyses using UV laser pulses are carried out on a commercial LAWATAP instrument (CAMECA Instruments). The UV pulses are generated using an amplified ytterbium-doped laser (S-pulse, Amplitude) delivering 350 fs pulses at 100 KHz. The laser wavelength of 1030 nm can be non-linearly converted to green (515 nm) and UV (343 nm). For our work, 343 nm is used. The APT vacuum chamber is the same as that used for THz-APT analyses, however the UV beam is focused on the sample using a lens with a 20 cm focal length. The LAWATAP system is equipped with a custom-designed microchannel plate/advanced delay line detector (MCP/aDLD) and improved capabilities in revealing multiple detection events[50]. Electrical pulse APT analyses were carried out on a commercial LEAP 5000 XS instrument (CAMECA Instruments). For all analyses, the 3D reconstruction of the evaporated volume is obtained using standard $V_{DC}$-driven reconstruction protocol adjusting the sample's radius according to the changes in voltage. The reconstructions parameters such as the image compression factor and field factor were determined using the crystallographic information of the samples deducted from the detector hit maps[51]. For consecutive measurements on the same specimen, acquisition was continued after each change in pulsing mode until stable evaporation conditions were reached, as assessed from the mass spectrum and hit map. Typically, several hundred thousand ions were evaporated during this stabilization period before the datasets presented here were recorded.

## ASSOCIATED CONTENT

**Supporting Information.**

The Supplementary material is available free of charge. It includes a PDF file with: details on the THz beam's spatial profile, details on the THz beam polarization, details on the electro-optic sampling method, a description of the electric field estimation method, details on the electron field emission experiment, details on the field and sample geometry temporal evolution, details on the APT samples and details on the APT data quality metrics.

AUTHOR INFORMATION

**Corresponding Author**

*angela.vella@univ-rouen.fr

*hideur@coria.fr

**Author Contributions**

L. G., F.E., A.F, and J.H. designed and conducted the experiments. L.G., F.E., M.D.T. and J.H. analyzed the results. S.I., J.H. and G.D.C contributed to the implementation of the setup. A.M., T.G., X.R., D.P. and F.B. gave conceptual suggestions. F.E., M.D.T. and J.H. provided the samples. A.V. and A.H. conceived and supervised the research. The original draft was written by L.G. and F.E. All authors reviewed the manuscript.

**Notes**

The authors declare no competing interests.

FUNDING SOURCES

The research leading to these results has received funding from the European Innovation Council under grant agreement No.101046651- MIMOSA and the Normandy region.

## ACKNOWLEDGMENT

We gratefully acknowledge Marie-Hélène Mammez et Jean-Francois Lampin for their assistance in the calibration of the thermopile THz detector used in this work.

## ABBREVIATIONS

aDLD, advanced delay line detector; APT, atom probe tomography; EO, electro-optic; FW0.1M, full width at ten maximum; FWHM, full width at half maximum; ITO, indium tin oxide; LN, lithium niobate; MCP, microchannel plates; MRP, mass resolving power; OAEM, off-axis ellipsoidal mirror; OAPM, off-axis parabolic mirror; OR, optical rectification; PSD, position sensitive detector; SNR, signal-to-noise ratio; TFP, thin film polarizer; THz, terahertz; TOF, time of flight; TPF, tilted pulse-front; VP, voltage pulsing; Yb, ytterbium;

SYNOPSIS

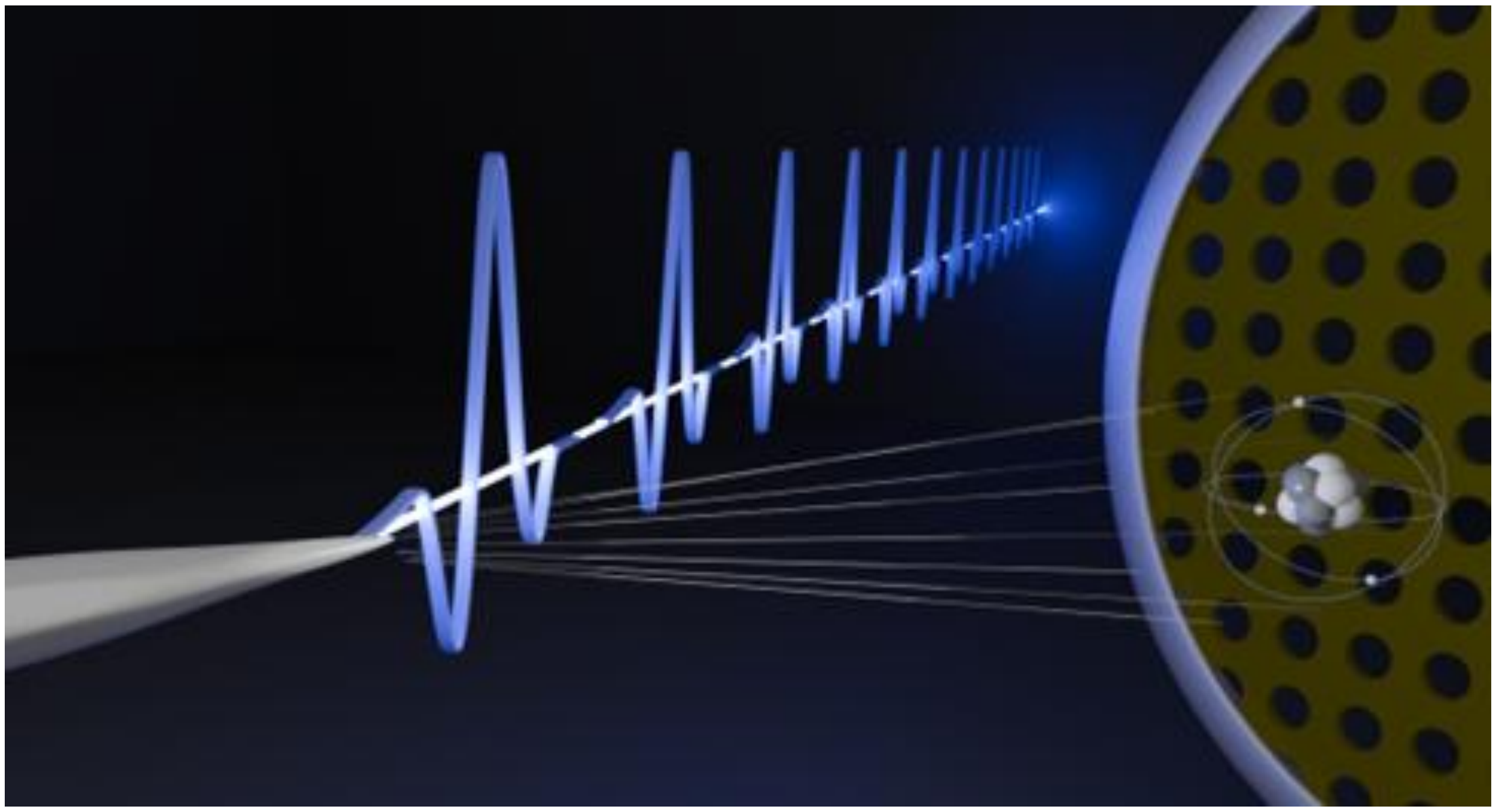